%% file: main.tex
\documentclass{article}

\usepackage[preprint]{neurips_2026}

\usepackage[utf8]{inputenc}
\usepackage[T1]{fontenc}
\usepackage{hyperref}
\usepackage{url}
\usepackage{booktabs}
\usepackage{float}
\usepackage{overpic}
\usepackage{wrapfig}
\usepackage{subcaption}
\usepackage{stfloats}
\usepackage{bm}
\usepackage{amsmath}
\usepackage{amsfonts}
\usepackage{multirow}
\usepackage{multicol}
\usepackage{enumitem}
\usepackage{threeparttable}
\usepackage{xspace}
\usepackage[normalem]{ulem}
\usepackage{algorithm}
\usepackage{algorithmicx}
\usepackage{algpseudocode}
\usepackage{marvosym}
\usepackage{colortbl}
\usepackage{tcolorbox}
\tcbuselibrary{breakable}
\usepackage{wasysym}
\usepackage[figuresright]{rotating}
\usepackage{pifont}
\usepackage{microtype}
\usepackage{xcolor}
\usepackage{nicefrac}

\newcommand{\methodshort}{\textsc{HD-Rec}}

\title{Hierarchical Quantization with Domain-Adaptive Sparse Routing for Generative Cross-Domain Recommendation}

\author{%
\textbf{
Haiying He$^{1}$\footnotemark[1],
Xiaopeng Li$^{1}$\footnotemark[1],
Yuchen Gu$^{1}$,
Kuo Cai$^{2}$,
Bo Chen$^{2}$,
Jingtong Gao$^{1}$,
Yejing Wang$^{1}$}\\
\textbf{Derong Xu$^{1}$,
Qiang Luo$^{2}$,
Ruiming Tang$^{2}$,
Guorui Zhou$^{2}$,
Han Li$^{2}$,
Xiangyu Zhao$^{1}$\footnotemark[2]}\\[2pt]
{\normalfont $^{1}$City University of Hong Kong},
{\normalfont $^{2}$Kuaishou Technology}%
}

\begin{document}

\maketitle

\begin{abstract}
Generative Recommendation (GenRec) represents a promising paradigm that achieves remarkable empirical success by encoding items as compact Semantic IDs (SIDs) and modeling user behavior via next-token prediction across diverse recommendation scenarios.
 Extending this
paradigm to cross-domain recommendation is challenging because a unified
model must accommodate heterogeneous item semantics and behavioral
patterns across domains. Existing methods commonly rely on globally shared
representations or lightweight domain adaptation, which may provide
insufficient capacity for modeling heterogeneous patterns at different
semantic granularities.
To address these challenges, we propose \methodshort{}, a unified generative framework for cross-domain
recommendation. \methodshort{} employs a hierarchical domain-aware quantizer that
constructs semantic identifiers using globally shared coarse-level
codebooks and adaptively routed fine-level codebooks. It further introduces
a domain-adaptive sparse mixture-of-experts module that combines a
continuously activated shared expert with a dynamically selected
specialized expert. To improve the coherence of multi-token item
representations, we develop a cross-granularity routing consistency
objective that regularizes token-level routing decisions toward their
item-level consensus.
Experiments on three public cross-domain recommendation benchmarks show
that \methodshort{} consistently improves over competitive sequential, generative,
and cross-domain recommendation baselines. 
\end{abstract}

\input{1Introduction}
\input{2Preliminary}

\input{3Method}
\input{4Experiment}

\input{5RelatedWork}
\input{6Conclusion}

\input{7Limitation}



\bibliographystyle{plainnat}
\bibliography{main}

\end{document}

%% file: 1Introduction.tex
\section{Introduction}  \label{sec:intro}

Recommender systems have become indispensable across e-commerce, online media, and social platforms~\cite{li2025li_survey}. Their primary goal is to assist users in efficiently discovering relevant items within vast information, thereby enhancing user experience and engagement. In practical deployments, they frequently operate across multiple domains or scenarios (e.g., distinct content categories)~\cite{zhu2021cross, zang2022survey_}. In such contexts, leveraging data from resource-rich domains can significantly improve recommendations in data-scarce domains, establishing the critical need for effective Cross-Domain Recommendation (CDR). Early CDR approaches primarily relied on overlapping entities (e.g., shared users or items) to align domains and facilitate knowledge transfer~\cite{li2011cross}. However, the applicability of these methods is restricted in practical scenarios where domains share minimal or zero common identifiers due to privacy constraints or platform heterogeneity~\cite{zhang2022cross}. Furthermore, even when entity overlap is present, manually engineered feature alignments frequently fail to capture the semantic relationships between domains possessing significantly different content and user bases. To overcome the limitation of relying on overlapping entities, recent methods have explored utilizing content features through pre-trained language models (PLMs) to construct unified semantic spaces~\cite{hou2022towards,geng2022recommendation}. These approaches encode item metadata (e.g., titles and descriptions) into transferable representations that bridge domains without explicit entity correspondence. Although promising, they typically necessitate extensive auxiliary data for effective pre-training and may encounter domain gap issues when content distributions differ substantially~\cite{su2024cross}. These constraints indicate that existing solutions for CDR have not fully utilized the semantics of cross-domain content in a unified and scalable manner. Consequently, there is a compelling motivation to explore a new recommendation paradigm capable of transferring knowledge across heterogeneous domains without the requirement for explicit entity overlaps.

\begin{wrapfigure}{r}{0.5\textwidth} 
  \centering
  \vspace{-0.6cm}
  \includegraphics[width=\linewidth]{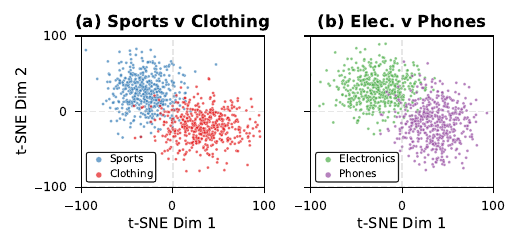}
  \caption{t-SNE visualization of the learned embeddings. (a) Comparison between Sports and Clothing domains. (b) Comparison between Electronics and Cell Phones domains.}
  \label{fig:tsne}
  \vspace{-0.5cm} 
\end{wrapfigure}

To overcome these limitations, recent studies explore Generative Recommendation (GenRec), a new paradigm reformulating recommendation as a sequence-to-sequence generation task~\cite{jin2025generative,deldjoo2024recommendation}. By encoding items into content-derived Semantic IDs (SIDs) and modeling user behavior via next-token prediction over item sequences, GenRec naturally unifies heterogeneous domains within a single generative framework. Building upon this, several studies extend GenRec to cross-domain settings. Some approaches map multi-domain items and interactions into a shared semantic space, training unified generative models over mixed-domain sequences~\cite{jin2025generative}. Others introduce additional components like domain prompts, lightweight adapters, or auxiliary contrastive losses to adapt models to domain-specific patterns~\cite{lee2025gram,chen2023cdr}. 
Nevertheless, 
a unified model must represent item embeddings that exhibit both cross-domain overlap and domain-dependent geometric structures.
To illustrate this challenge, we visualized item embeddings from the Amazon dataset using t-SNE (Figure~\ref{fig:tsne}). 
This visualization serves as a qualitative motivation rather than a quantitative assessment of representation disentanglement: a unified tokenizer needs shared capacity for common semantic structures while retaining adaptive capacity for heterogeneous patterns.

This observation motivates two design considerations for unified
generative cross-domain recommendation.
First, hierarchical semantic organization should represent coarse item semantics with globally shared capacity while allowing fine-grained patterns to be modeled adaptively. Second, adaptive capacity allocation should enable a unified model to activate specialized parameters according to item representations, rather than applying exactly the same parameters to all domains and items.

To address these challenges, we propose \methodshort{},
a unified generative framework for cross-domain recommendation.
Our core philosophy is hierarchical modeling: \methodshort{} first constructs multi-level semantic identifiers through hierarchical quantization: coarse-level codes are selected from globally shared codebooks, while the final residual code is selected from a set of adaptively routed codebooks. It then introduces a domain-adaptive sparse mixture-of-experts module that combines an always-active shared expert with a dynamically selected specialized expert. Finally, cross-granularity routing consistency learning encourages the semantic tokens belonging to the same item to maintain coherent routing decisions. Together, these components provide hierarchical representation capacity and sparse conditional computation for unified cross-domain generative modeling.
To summarize, the main contributions are as follows:

\begin{itemize}[leftmargin=*]
    \item We propose \methodshort{}, a unified generative framework for cross-domain
  recommendation that combines hierarchical semantic tokenization with
  domain-adaptive sparse routing.

 \item We introduce a hierarchical domain-aware quantizer that combines globally shared coarse-level codebooks with adaptively routed fine-level codebooks, together with a sparse MoE module that combines shared and specialized model capacity.

 \item We develop a cross-granularity routing consistency objective that
  encourages coherent expert-routing decisions among semantic tokens
  belonging to the same item.
  
    \item Extensive experiments on three cross-domain datasets show \methodshort{} significantly outperforms competitive baselines, validating its effectiveness.
\end{itemize}

%% file: 2Preliminary.tex
\section{Problem Definition}

In this section, we present the formal definition of the Cross-Domain Sequential Recommendation (CDSR) task. Consider a recommendation system consisting of a domain set $\mathcal{D} = \{d_1, d_2, \ldots, d_T\}$ and a user set $\mathcal{U}$. Each domain $d \in \mathcal{D}$ is associated with an independent item set $\mathcal{I}_d$, and the overall item space is $\mathcal{I} = \bigcup_{d_i \in \mathcal{D}} \mathcal{I}_{d_i}$.

\textbf{User Behavior Modeling.} 
For each user $u \in \mathcal{U}$, interactions within domain $d$ are chronologically ordered as:
\begin{equation}
S_u^d = (v_1, v_2, \ldots, v_{|S_u^d|}), \quad v_i \in \mathcal{I}_d.
\end{equation} 
Let $\mathcal{D}_u \subseteq \mathcal{D}$ denote the set of domains that user $u$ has interacted with. 
The full multi-domain behavior of user $u$ is defined as:
\begin{equation}
\mathcal{H}^u = \{S_{d_i}^u \mid d_i \in \mathcal{D}_u\}.
\end{equation}

\textbf{Cross-Domain Recommendation Task (CDSR).} 
Given a user $u$ and a target domain $d_t \in \mathcal{D}$, CDSR aims to predict the next item interaction in the target domain, conditioned on the user’s multi-domain behavior history $H_u$. 

Formally, for each target-domain interaction event at time $t$, we predict:
\begin{equation}
v_t \in \mathcal{I}_{d_t},
\end{equation}
based on historical behavior $H_u^{<t}$.

We model the probability of the next interaction using a parameterized model $P_\theta$:
\begin{equation}
P_\theta(v_t \mid H_u^{<t}, d_t).
\end{equation}

The model is optimized by maximizing the log-likelihood over all target-domain interactions:
\begin{equation}
\mathcal{L}(\theta)
=
\sum_{u \in \mathcal{U}}
\sum_{t \in \mathcal{T}_u^{d_t}}
\log P_\theta(v_t \mid H_u^{<t}, d_t),
\end{equation}

where $\mathcal{T}_u^{d_t}$ denotes the set of timestamps where user $u$ has interactions in the target domain $d_t$.

The formulation permits more than two domains. In this work, we focus on the case of two domains, i.e.,
$\mathcal{D} = \{d_s, d_t\}$.

%% file: 3Method.tex
\section{Methodology}

\begin{figure*}[t]
\centering
\includegraphics[width=\textwidth]{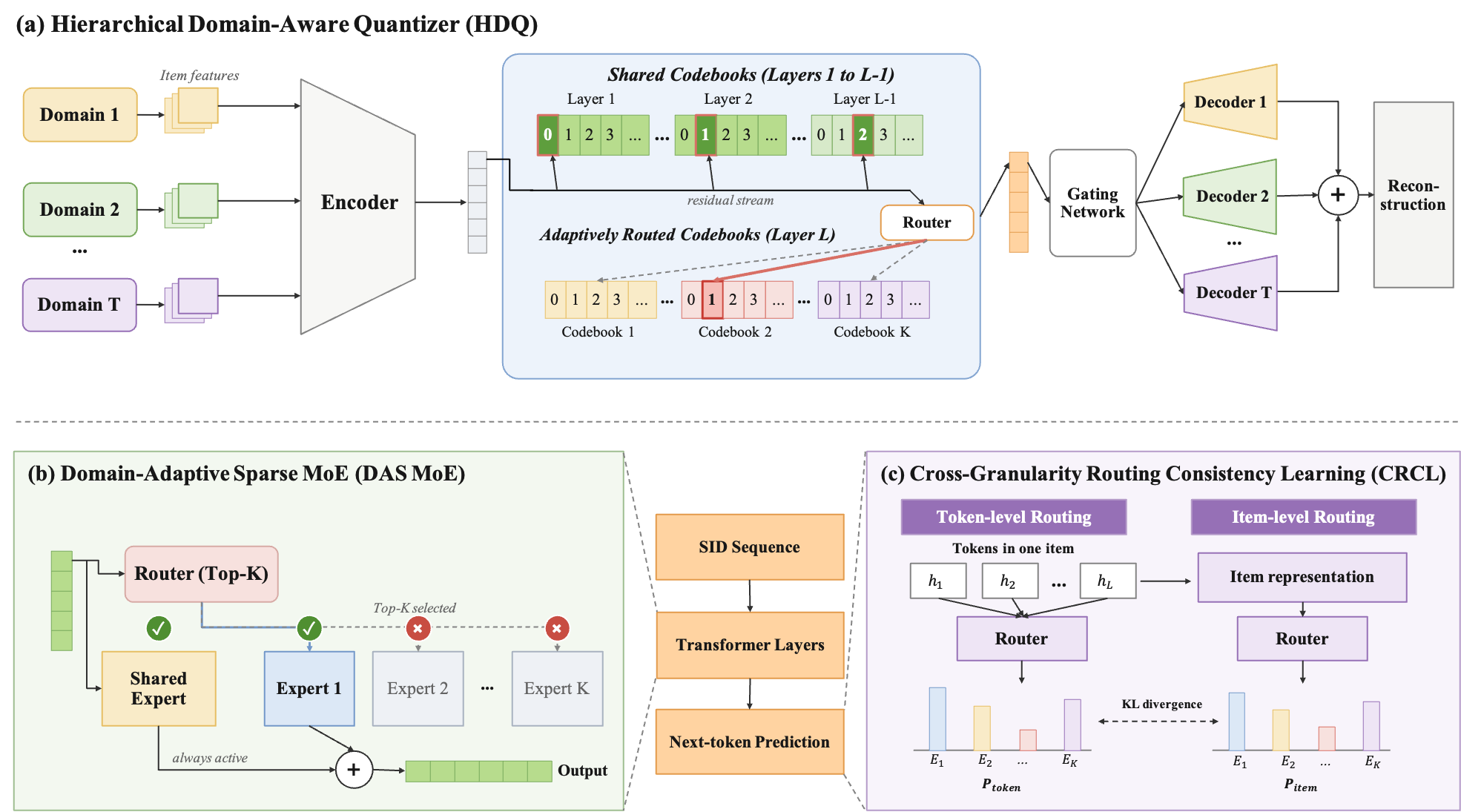}
\caption{The overview of the proposed \methodshort{}. (a) HDQ constructs Semantic IDs through globally shared coarse-level quantization and adaptively routed fine-level quantization. (b) DAS MoE combines an always-active shared expert with a dynamically selected specialized expert. (c) CRCL encourages coherent expert-routing decisions among tokens belonging to the same item.}
\label{fig:framework}
\vspace{-2mm}
\end{figure*}

In this section, we present details of \methodshort{}. As illustrated in Figure~\ref{fig:framework}, \methodshort{} consists of three key components: (i) a Hierarchical Domain-Aware Quantizer that constructs multi-level Semantic IDs using globally shared coarse-level codebooks and adaptively routed fine-level codebooks (Section~\ref{sec:tokenizer}); (ii) a Domain-Adaptive Sparse MoE that combines an always-active shared expert with a dynamically selected specialized expert (Section~\ref{sec:moe}); and (iii) a Cross-Granularity Routing Consistency Learning that regularizes token-level routing decisions toward their item-level consensus (Section~\ref{sec:loss}). We elaborate on each component below.

\subsection{Hierarchical Domain-aware Quantizer} 
\label{sec:tokenizer}
Conventional residual quantization methods such as Residual-Quantized Variational Autoencoder (RQ-VAE)~\cite{lee2022autoregressive} learn shared codebooks across all domains.
While effective for representation compression, 
this design may under represent domain-dependent variations in cross-domain recommendation.
In cross-domain recommendation, however, items often share high-level semantics (e.g., ``sports electronics'', ``basketball'') while exhibiting specific variations in fine-grained preferences (e.g., “low-top basketball shoes” in e-commerce versus “basketball training videos” on content platforms).


To address this limitation, we propose a \textbf{Hierarchical Domain-Aware Quantizer} (HDQ) that progressively decomposes item semantics. 
Our core insight is to introduce a hierarchical inductive bias for organizing item semantics: the shared shallow layers model coarse semantic structures, whereas the
routed final layer provides additional capacity for fine-grained residual
quantization.

\textbf{Coarse-Grained Semantic Decomposition (Layers $1$ to $L-1$).} 
Given an item $i$ with a continuous feature vector $\mathbf{x}_i \in \mathbb{R}^D$, an encoder $E_\theta$ maps it to a latent representation $\mathbf{z}_i = E_\theta(\mathbf{x}_i) \in \mathbb{R}^d$. We initialize the residual as $\mathbf{r}_0 = \mathbf{z}_i$. For the first $L-1$ layers, the tokenizer progressively captures coarse-level semantics using globally shared codebooks. At each layer $l \in \{1, \ldots, L-1\}$, it selects the nearest code vector from the shared codebook $\mathcal{C}_l \subset \mathbb{R}^d$ by minimizing the Euclidean distance:
\begin{align}
\mathbf{c}_l = \underset{\mathbf{c} \in \mathcal{C}_l}{\arg\min} \|\mathbf{r}_{l-1} - \mathbf{c}\|_2 
\end{align}
The residual is subsequently updated by subtracting the selected vector to propagate the remaining information to the next layer:
\begin{align}
\mathbf{r}_l = \mathbf{r}_{l-1} - \mathbf{c}_l
\end{align}

\textbf{Domain-Adaptive Routing and Specialization (Layer $L$).}
Unlike the preceding layers, the final quantization layer $L$ 
provides adaptive fine-grained quantization.
We introduce a specialized codebook decomposition strategy comprising $K$ distinct codebooks $\{\mathcal{C}_L^{(1)}, \ldots, \mathcal{C}_L^{(K)}\}$, 
each providing additional capacity for residual representation. Rather than assigning these codebooks to predefined domains, a lightweight router dynamically selects a codebook according to the item's residual representation.
Given the final residual $\mathbf{r}_{L-1}$, the router $g_\phi$ yields routing logits over the $K$ candidate codebooks:
\begin{align}
    \boldsymbol{\alpha}_i = g_\phi(\mathbf{r}_{L-1}) \in \mathbb{R}^K,
\end{align}
To obtain a differentiable, discrete routing decision, we apply the Gumbel-Softmax trick~\cite{jang2016categorical}:
\begin{align}
    \mathbf{a}_i = \mathrm{GumbelSoftmax}(\boldsymbol{\alpha}_i, \tau) \in \mathbb{R}^K,
\end{align}
where $\tau$ is the temperature parameter. During the forward pass, we adopt a straight-through estimator to deterministically select the top-1 codebook index $k_i = \arg\max_{k} a_{ik}$, while utilizing the soft Gumbel-Softmax probabilities during the backward pass for gradient propagation. 
Thus, the learned routing index $k_i$ dynamically assigns the item to an adaptive semantic subspace. The tokenizer then performs a nearest-neighbor lookup exclusively within the selected codebook:
\begin{equation}
    \mathbf{c}_L^{(k_i)} = \underset{\mathbf{c} \in \mathcal{C}_L^{(k_i)}}{\arg\min} \|\mathbf{r}_{L-1} - \mathbf{c}\|_2,
\end{equation}
where $\mathbf{r}_{L-1}$ denotes the residual propagated from layer $L-1$. Each routed codebook is assigned a unique offset to ensure that its codewords have distinct Semantic IDs in the global vocabulary.

\textbf{Final Representation and Reconstruction.}
The final quantized latent representation $\hat{\mathbf{z}}_i$ is aggregated as the sum of the shared-codebook components from shallow layers and the routed adaptive component from the final layer:
\begin{equation}
    \hat{\mathbf{z}}_i = \sum_{l=1}^{L-1} \mathbf{c}_l + \mathbf{c}_L^{(k_i)}.
\end{equation}
A decoder $D_\psi$ is then employed to reconstruct the original feature from this quantized representation, yielding $\hat{\mathbf{x}}_i = D_\psi(\hat{\mathbf{z}}_i)$.

\textbf{Joint Optimization Objectives.}
The hierarchical quantizer is trained end-to-end. We use the mean squared reconstruction error over a mini-batch $\mathcal{B}$:
To prevent routing collapse, a common degenerate solution where only a few codebooks receive the majority of assignments, we introduce a load-balancing regularizer. Let $u_k = \frac{1}{|\mathcal{B}|} \sum_{i \in \mathcal{B}} a_{ik}$ denote the average routing probability of codebook $k$ within the mini-batch. We encourage uniform utilization across all $K$ codebooks:
\begin{align}
\mathcal{L}_{\text{balance}} = \sum_{k=1}^{K} \left(u_k - \frac{1}{K}\right)^2
\end{align}
Finally, to update the codebooks while stabilizing the encoder training, we apply the standard vector quantization losses using the stop-gradient operator $\mathrm{sg}[\cdot]$. The codebook loss ($\mathcal{L}_{\text{cb}}$) and commitment loss ($\mathcal{L}_{\text{com}}$) are applied across all shared layers and the routed final layer:
\begin{align}
    \mathcal{L}_{\text{cb}}
    &= \frac{1}{|\mathcal{B}|} \sum_{i \in \mathcal{B}}
    \Bigg(
    \sum_{l=1}^{L-1}
    \left\| \mathrm{sg}[\mathbf{r}_{l-1}^{(i)}] - \mathbf{c}_l^{(i)} \right\|_2^2
    + \left\| \mathrm{sg}[\mathbf{r}_{L-1}^{(i)}] - \mathbf{c}_L^{(k_i)} \right\|_2^2
    \Bigg),\\
    \mathcal{L}_{\text{com}}
    &= \frac{1}{|\mathcal{B}|} \sum_{i \in \mathcal{B}}
    \Bigg(
    \sum_{l=1}^{L-1}
    \left\| \mathbf{r}_{l-1}^{(i)} - \mathrm{sg}[\mathbf{c}_l^{(i)}] \right\|_2^2
    + \left\| \mathbf{r}_{L-1}^{(i)} - \mathrm{sg}[\mathbf{c}_L^{(k_i)}] \right\|_2^2
    \Bigg),
\end{align}
The overall training objective is formulated as the weighted sum of these components:
\begin{align}
    \mathcal{L}
    = \mathcal{L}_{\text{rec}}
    + \lambda_{\text{cb}} \mathcal{L}_{\text{cb}}
    + \lambda_{\text{com}} \mathcal{L}_{\text{com}}
    + \lambda_{\text{bal}} \mathcal{L}_{\text{balance}}.
\end{align}

\subsection{Domain-Adaptive Sparse MoE}
\label{sec:moe}
While hierarchical quantization organizes item representations at the input level, a standard Transformer backbone applies the same dense FFN parameters to all tokens. 
To provide conditional model capacity under unified cross-domain training,
we introduce a \textbf{Domain-Adaptive Sparse Mixture-of-Experts (DAS MoE)} architecture. 
The module combines an always-active shared expert with one input-selected specialized expert, thereby allocating additional capacity in a sparse and data-dependent manner.

Specifically, our MoE layer comprises a single shared expert $E_{\text{shared}}$ and $K$ specialized experts $\{E_1, \ldots, E_K\}$, where each expert is instantiated as a Feed-Forward Network (FFN). For a token embedding $\mathbf{h}_t \in \mathbb{R}^d$ at position $t$, a lightweight gating network projects the representation to compute routing logits over the $K$ specialized experts:
\begin{align}
s_t=W_{gate}\mathbf{h}_t \in \mathbb{R}^K
\end{align}
where $W_{\text{gate}}$ denotes the learnable parameters of the router. Same to the HDQ router, We apply a sparse routing mechanism to deterministically select the top-1 specialized-expert routing index $k_t^*$:
\begin{align}
k_t^* = 
    \underset{k}{\arg\max} s_{t,k},
\end{align}
The final token representation combines the outputs of the always-active shared expert and the selected specialized expert:
\begin{align}
    MoE(h_t)=E_{shared}(\mathbf{h}_t)+E_{k_t^*}(\mathbf{h}_t).
\end{align}

This architecture offers several critical advantages. 
Specifically, the shared expert processes every token, while the routing network selects one additional expert based on the token representation. The resulting shared-plus-routed structure provides conditional model capacity while preserving sparse computation, since only one routed expert is activated for each token.


\subsection{Cross-granularity Routing Consistency Learning}
\label{sec:loss}
In generative recommendation, a cross-domain interaction sequence for a user can be represented as $\boldsymbol{S}_u = [i_1, \ldots, i_n]$. Utilizing our hierarchical tokenizer, each item is decomposed into a sequence of $L$ semantic tokens, denoted as $[\boldsymbol{c}_1, \ldots, \boldsymbol{c}_L]$. The primary training objective of the backbone model is the standard next-token prediction (NTP):
\begin{equation}
\mathcal{L}_{ntp} = -\sum_{t=1}^{n-1}\sum_{l=1}^L{\log}P(\boldsymbol{c}_l^{(t+1)}|\boldsymbol{c}_{<l}^{(t+1)}, i_1,i_2, \dots, i_t)
\end{equation}
Standard MoE architectures typically operate strictly at the token level, where each token independently selects experts based on its local hidden representation~\cite{dai2024deepseekmoe}. While this design ensures maximum flexibility, it poses a unique challenge in our setting: 
Since the tokens of the same item jointly constitute its Semantic ID, unconstrained token-wise routing may assign them substantially different expert distributions, increasing intra-item routing variability.

To resolve this issue, we propose a soft \textbf{Cross-Granularity Routing Consistency regularization} (CRCL). This mechanism encourages routing coherence among tokens belonging to the same item by minimizing the Kullback-Leibler (KL) divergence between each individual token’s routing distribution and the aggregated item-level distribution. 
This regularizer encourages tokens belonging to the same item to exhibit similar routing distributions while retaining token-level routing flexibility.

Specifically, for an item $i$ consisting of $L$ token embeddings $\{\boldsymbol{h}_1^{(i)}, \ldots, \mathbf{h}_L^{(i)}\}$, we first construct an item-level global representation to serve as an item-level routing reference obtained by mean pooling its token embeddings:
\begin{align}
    \mathbf{h}_{avg} = \frac{1}{L}\sum_{t=1}^{L}\mathbf{h}_t^{(i)}
\end{align}
We then project this global representation through the gating network to obtain the item-level routing logits $s_i = W_{\text{gate}}\mathbf{h}_{\text{avg}}^{(i)} \in \mathbb{R}^K$, which induce an item-level routing distribution:
\begin{align}
P_{item} = softmax(s_i).
\end{align}
Concurrently, each individual token $t$ maintains its own localized routing distribution based on its specific embedding:
\begin{align}
P_{\text{token}}^{(t)} = \text{softmax}(W_{\text{gate}}\mathbf{h}_t^{(i)}). 
\end{align}
To enforce consistency, the Cross-Granularity Routing Consistency Loss penalizes the divergence of token-level routing decisions from the consensus item-level distribution:
\begin{align}
    \mathcal{L}_{consistency}= \frac{1}{|\mathcal{B}|} \sum_{i\in\mathcal{B}}\frac{1}{L}\sum_{t=1}^L\mathbf{KL}(P_{token}\|P_{item}),
\end{align}
where $\mathcal{B}$ denotes the mini-batch. 
This regularization encourages tokens belonging to the same item to maintain similar routing preferences while preserving token-level routing flexibility through the soft KL-divergence objective.
The final optimization objective is formulated by jointly minimizing the generative task loss, the consistency regularization, and the MoE load-balancing loss:
\begin{equation}
    \mathcal{L}_{total} = \mathcal{L}_{ntp} + \lambda_1\mathcal{L}_{consistency}+\lambda_2\mathcal{L}_{balance}
\end{equation}
where $\lambda_1, \lambda_2 \in [0, 1]$ are hyper-parameters carefully tuned to control the strength of the respective regularizations.


%% file: 4Experiment.tex
\section{Experiment}


In this section, we conduct extensive experiments on several publicly available datasets to evaluate the effectiveness of our proposed method. To systematically analyze the performance of \methodshort{}, we investigate several \textbf{Research Questions} (\textbf{RQ}) based on the experimental results.

\begin{itemize}[leftmargin=*]
    \item \textbf{RQ1}: How does \methodshort{} perform compared with existing cross-domain and single-domain recommendation models?
    \item \textbf{RQ2}: How does each component contribute to the overall performance of \methodshort{}?
    \item \textbf{RQ3}: How do hyperparameters affect the performance of \methodshort{}?
    \item \textbf{RQ4}: Does the proposed architecture introduce significant inference overhead?
    \item \textbf{RQ5}:
    Does \methodshort{} improve intra-item routing consistency?

\end{itemize}

\subsection{Experimental Setup}
\subsubsection{{\textbf{Datasets}}}   \label{sec:exp-data}
We conduct experiments on three public cross-domain dataset pairs: \textbf{Clothing-Sports} (Leisure), \textbf{Electronics-Phones} (Technology) and \textbf{Books-Movies} (Entertainment). The first two datasets are constructed from sub-categories of the Amazon product review dataset\footnote{https://jmcauley.ucsd.edu/data/amazon/index\_2014.html/}, which contains user reviews of products on a large e-commerce platform. The third  is derived from the Douban dataset\footnote{\url{https://www.researchgate.net/publication/350793434_Douban_dataset_ratings_item_details_user_profiles_tags_and_reviews}}, which contains user reviews of books, movies, and music on a major Chinese social platform.

For model evaluation, we adopt the leave-last-out evaluation protocol. For each user interaction sequence, the second-to-last interaction is used for validation and the last one is used for testing. 
After preprocessing, the statistics of the dataset are summarized in Table~\ref{tab:exp-dataset}.

\begin{wraptable}{r}{0.45\textwidth}
\vspace{-2mm}
\centering
\scriptsize
\setlength{\tabcolsep}{2.4pt}
\vspace{-1cm}
\caption{Dataset statistics.}
\vspace{-2mm}
\begin{tabular}{lrrrr}
\toprule
\textbf{Data} & \textbf{Users} & \textbf{Items} & \textbf{Inter.} & \textbf{Spar.} \\
\midrule
Clothing & 39,387 & 23,033 & 278,677 & 99.97\% \\
Sports & 35,598 & 18,357 & 296,337 & 99.95\% \\
\midrule
Electronics & 192,403 & 63,001 & 1,689,188 & 99.99\% \\
Phones & 27,879 & 10,429 & 194,439 & 99.93\% \\
\midrule
Books & 1,713 & 8,601 & 104,295 & 99.29\% \\
Movies & 2,628 & 20,964 & 1,249,016 & 97.73\% \\
\bottomrule
\end{tabular}
\label{tab:exp-dataset}
\vspace{-5mm}
\end{wraptable}

\subsubsection{\textbf{Baselines}}
To comprehensively validate the effectiveness of \methodshort{}, We benchmark our approach against three major categories of state-of-the-art baselines:

\noindent \textbf{i) Single-domain Sequential Recommendation} (SDSR). These methods model users’ sequential behaviors within an individual domain using advanced neural architectures or sparsity-aware training strategies. Due to their single-domain setting, these methods do not leverage cross-domain signals or auxiliary semantic information.
\begin{itemize}[leftmargin=*]
    \item \textbf{BERT4Rec}~\cite{sun2019bert4rec} adapts the Transformer-based BERT~\cite{kenton2019bert} framework to sequential recommendation by introducing a masked item prediction objective. 
    \item \textbf{SASRec}~\cite{kang2018self} adopts a unidirectional self-attention mechanism to model user behavior sequences, thereby preserving the temporal causality in sequential interactions. 
    \item \textbf{STOSA}~\cite{fan2022sequential} enriches the self-attention mechanism by modeling item representations as distributions and adopting Wasserstein-based attention to capture collaborative transitivity.
\end{itemize}

    \noindent \textbf{ii) Generative Recommendation Systems} (GenRec). This line of work formulates recommendation as an autoregressive generation task, typically leveraging large pre-trained language models or generative architectures to generate item identifiers or semantic tokens based on user interaction sequences. However, most existing approaches are developed under a single-domain setting and primarily rely on in-domain interaction data, without explicitly modeling transferable knowledge across domains.

\begin{itemize}[leftmargin=*]
    \item  \textbf{VQ-Rec}~\cite{hou2023learning} discretizes continuous item embeddings into codebook indices via vector-quantized tokenization and performs prediction in the discrete token space using Transformers.
    
    \item  \textbf{TIGER}~\cite{rajput2023recommender} represents each item as a semantic identifier composed of quantized codewords derived from content embeddings and trains a Transformer to generate next‑item Semantic IDs.
    
    \item  \textbf{HSTU}~\cite{zhai2024actions} formulates recommendation as autoregressive transduction and uses hierarchical self-attention to encode long behavioral histories into unified sequential representations for joint ranking and retrieval generation.

\end{itemize}

\noindent \textbf{iii) Cross-domain Sequential Recommendation} (CDSR). These methods leverage user interactions across multiple domains to mitigate data sparsity and improve sequential recommendation performance. By transferring knowledge across domains, these approaches can capture richer user preference patterns than single-domain models.
\begin{itemize}[leftmargin=*]
    \item  \textbf{C2DSR}~\cite{cao2022contrastive_} constructs a cross-domain interaction graph and applies graph neural networks with adaptive gating mechanism to regulate knowledge transfer.
    
    \item  \textbf{TriCDR}~\cite{ma2024triple} employs triplet-based contrastive learning to align users embeddings across domains and enhances cross-domain preference modeling.
    \item  \textbf{LLM4CDSR}~\cite{liu2025bridge} reformulates CDSR as a text generation task by converting interaction histories into textual prompts via semantic embeddings and hierarchical user profiling to capture implicit cross-domain preferences.

    \item  \textbf{GenCDR}~\cite{hu2025ids} introduces a generative cross-domain framework with a domain-adaptive tokenizer (universal RQ-VAE + domain-specific LoRA~\cite{hu2022lora}) and a domain-adaptive recommender (universal LoRA on aggregated data + domain-specific LoRA), both fine-tuned via parameter-efficient LoRA. 
    In contrast, \methodshort{} introduces input-dependent capacity at both the tokenizer and backbone levels through routed fine-level codebooks and sparse experts. \methodshort{} further regularizes routing consistency across the semantic tokens of the same item. The two approaches therefore explore different parameterizations for unified generative cross-domain recommendation.
\end{itemize}

\subsubsection{\textbf{Implementation Details}}

All experiments are conducted on NVIDIA A800 GPUs. \methodshort{} is trained using a two-stage pipeline. In the first stage, we train the hierarchical domain-aware tokenizer on all item embeddings. We use AdamW as the optimizer with a learning rate of 1e-3. The group size of each codebook is set to 1024 and the code dimension is set to 32. The routing network is implemented as a single-layer MLP with ReLU activation. In the second stage, we train the entire recommendation model. We use T5 ~\cite{raffel2020exploring} as the backbone architecture and set the learning rate to 1e-4. During inference for the main performance evaluation, beam search with beam size 20 is applied. 
For the efficiency analysis in RQ4, we follow a stress-test setting with beam size 200 to evaluate the inference overhead under a larger candidate search space. The weight of the Cross-granularity Routing Consistency loss is set to 0.01.

\subsubsection{\textbf{Evaluation Metrics}}
Following the standard practice in previous works~\cite{kang2018self,cao2022contrastive_}, we evaluate model performance using standard Top-k recommendation metrics, including \textit{\textbf{Hit Rate@10}} and \textit{\textbf{NDCG@10}}. Model selection is conducted by choosing the checkpoint that achieves the highest \textit{\textbf{Hit Rate@10}} on the validation set, and the selected model is then evaluated on the test set.

\subsection{Overall Performance (RQ1)}
Table~\ref{tab:exp-overall} reports the overall performance of \methodshort{}. 
The results show that HD-REC performs favorably under unified
cross-domain training. In particular, the improvements are more
pronounced on Sports and Electronics, while smaller but consistent gains
are observed on Clothing and the Douban domains.

(1) \textbf{Single-domain Sequential Recommendation (SDSR)} models operate within a single domain and rely primarily on ID-based sequence signals. Since they neither exploit interactions from other domains nor leverage semantic content information, they consistently exhibit the weakest performance across domains, with particularly large gaps on the Amazon sub-domain pairs.

(2) \textbf{Generative Recommendation Systems (GenRec)} approaches adopt a generative paradigm that models recommendation as an autoregressive prediction problem over semantic item tokens. By jointly training on mixed-domain sequences and incorporating semantic representations,
these methods alleviate the limitations of pure ID-based models and achieve consistent improvements over SDSR. 
However, these methods do not explicitly introduce hierarchical adaptive capacity for heterogeneous item representations, which may limit their flexibility under unified cross-domain training.


(3) \textbf{Cross-domain Sequential Recommendation (CDSR)} baselines further improve performance by explicitly modeling cross-domain interactions through alignment mechanisms, auxiliary domain information, or LLM-based semantic bridging. Consequently, they generally outperform GenRec models, highlighting the effectiveness of leveraging cross-domain signals to mitigate data sparsity. Nevertheless, most CDSR methods still rely on shared representation spaces or relatively simple transfer mechanisms, their representation and adaptation mechanisms differ from the hierarchical quantization and token-level sparse routing explored in this work.

(4) \textbf{\methodshort{}} achieves the best performance across all domains and metrics. 
The gains are particularly pronounced on Sports and Electronics, which are also among the sparser domains in the evaluated benchmarks.
Compared to the strongest baseline, \methodshort{} improves H@10 by 17.6\% on Sports, 16.3\% on Electronics, and 9.9\% on Phones, while maintaining consistent improvements on Clothing and Douban datasets. 
These results establish the empirical effectiveness of the proposed architecture under pairwise cross-domain training.

\begin{table*}[!t]
\tabcolsep=0.1cm 
\centering
\caption{The overall results of competing baselines and \methodshort{} on three cross-domain datasets. The best and second-best results are highlighted in \textbf{bold} and \underline{underline}, respectively. ``\textbf{Impr.}'' denotes the relative improvement of \methodshort{} over the strongest baseline. The t-tests showed significant performance improvements ($p \leq 0.05$).}
\resizebox{1\textwidth}{!}{
\begin{tabular}{cc|cccc|cccc|cccc}
\toprule
\multicolumn{2}{c|}{\multirow{3}{*}{\textbf{Model}}} & \multicolumn{4}{c|}{\textbf{Amazon} (Leisure)} & \multicolumn{4}{c|}{\textbf{Amazon} (Technology)} & \multicolumn{4}{c}{\textbf{Douban} (Entertainment)} \\ 
\cmidrule{3-14} 
\multicolumn{2}{c|}{} & \multicolumn{2}{c}{\textbf{Clothing}} & \multicolumn{2}{c|}{\textbf{Sports}} & \multicolumn{2}{c}{\textbf{Electronics}} & \multicolumn{2}{c|}{\textbf{Phones}} & \multicolumn{2}{c}{\textbf{Books}} & \multicolumn{2}{c}{\textbf{Movies}} \\ 
\cmidrule{3-14} 
\multicolumn{2}{c|}{} & \textbf{H@10} & \textbf{N@10} & \textbf{H@10} & \textbf{N@10} & \textbf{H@10} & \textbf{N@10} & \textbf{H@10} & \textbf{N@10} & \textbf{H@10} & \textbf{N@10} & \textbf{H@10} & \textbf{N@10} \\ 
\midrule
\multirow{4}{*}{\textbf{SDSR}} 
& BERT4Rec & 0.0219 & 0.0105 & 0.0325 & 0.0169 & 0.0276 & 0.0149 & 0.0524 & 0.0278 & 0.0176 & 0.0158 & 0.1798 & 0.1211 \\
& SASRec & 0.0227 & 0.0108 & 0.0334 &  0.0173 &  0.0285 &  0.0154 & 0.0537 & 0.0287 &  0.0182 & 0.0164 & 0.1825 &  0.1265 \\
& STOSA & 0.0223 & 0.0135 &  0.0346 & 0.0283 & 0.0315 & 0.0172 & 0.0618 &  0.0346 & 0.0219 &  0.0165 & 0.1753 & 0.1223 \\ 
\midrule
\multirow{3}{*}{\textbf{GenRec}} 
& VQ-Rec & 0.0248 & 0.0170 & 0.0389 & 0.0281 & 0.0318 & 0.0262 & 0.0607 & 0.0399 & 0.0224 & 0.0201 & 0.1922 & 0.1261 \\
& TIGER &  0.0241 & 0.0167 & 0.0397 & \underline{0.0287} & 0.0322 & 0.0269 & 0.0613 & 0.0406 & 0.0221 & 0.0198 & 0.1893 & 0.1255 \\
& HSTU & 0.0253 & 0.0174 & 0.0381 & 0.0277 & 0.0328 & 0.0271 & 0.0615 & 0.0425 & 0.0230 & 0.0206 & 0.1931 & 0.1268 \\
\midrule

\multirow{4}{*}{\textbf{CDSR}} 
& C2DSR & 0.0255 & 0.0191 & 0.0395 & 0.0258 & 0.0336 & 0.0278 & 0.0589 & 0.0493 & 0.0205 & 0.0182 & 0.1854 & 0.1203 \\
& TriCDR & 0.0258 & 0.0194 & 0.0396 & 0.0259 & 0.0339 & 0.0280 & 0.0593 & 0.0505 & 0.0211 & 0.0185 & 0.1865 & 0.1217 \\
& LLM4CDSR & 0.0261 & 0.0196 & 0.0398 & 0.0260 & 0.0338 & 0.0279 & 0.0614 & 0.0506 & 0.0216 & 0.0189 & 0.1878 & 0.1225 \\
& GenCDR & \underline{0.0265} & \underline{0.0203} & \underline{0.0403} & 0.0262 & \underline{0.0342} & \underline{0.0283} & \underline{0.0621} & \underline{0.0512} & \underline{0.0237} & \underline{0.0212} & \underline{0.1971} & \underline{0.1275} \\ 
\midrule

\multirow{2}{*}{\textbf{Ours}} 
& \cellcolor{gray!20}\textbf{\methodshort{}} & \cellcolor{gray!20}\textbf{0.0278}* & \cellcolor{gray!20}\textbf{0.0209}* & \cellcolor{gray!20}\textbf{0.0489}* & \cellcolor{gray!20}\textbf{0.0292}* & \cellcolor{gray!20}\textbf{0.0398}* & \cellcolor{gray!20}\textbf{0.0285}* & \cellcolor{gray!20}\textbf{0.0683}* & \cellcolor{gray!20}\textbf{0.0526}* & \cellcolor{gray!20}\textbf{0.0245}* & \cellcolor{gray!20}\textbf{0.0220}* & \cellcolor{gray!20}\textbf{0.1978}* & \cellcolor{gray!20}\textbf{0.1285}* \\
& Impr (\%) & 4.9\% & 2.9\% & 17.6\% & 1.7\% & 16.3\% & 0.7\% & 9.9\% & 2.7\% & 3.4\% & 3.6\% & 0.4\% & 0.8\% \\ 
\bottomrule
\end{tabular}
}
\label{tab:exp-overall}
\vspace{-3mm}
\end{table*}

\begin{wraptable}{r}{0.48\textwidth}
\vspace{-5mm}
\centering
\footnotesize
\setlength{\tabcolsep}{2.6pt}
\caption{Main ablation results.}
\vspace{-2mm}
\begin{tabular}{lcccc}
\toprule
\multirow{2}{*}{\textbf{Model}} & \multicolumn{2}{c}{\textbf{Cloth.}} & \multicolumn{2}{c}{\textbf{Sport.}} \\
\cmidrule{2-5}
& \textbf{H@10} & \textbf{N@10} & \textbf{H@10} & \textbf{N@10} \\
\midrule
\cellcolor{gray!20}\textbf{\methodshort{}} & \cellcolor{gray!20}\textbf{0.0278} & \cellcolor{gray!20}\textbf{0.0209} & \cellcolor{gray!20}\textbf{0.0489} & \cellcolor{gray!20}\textbf{0.0292} \\
\textit{w/o} HDQ & 0.0254 & 0.0188 & 0.0435 & 0.0261 \\
\textit{w/o} DAS MoE & 0.0256 & 0.0191 & 0.0418 & 0.0259 \\
\textit{w/o} CRCL & 0.0261 & 0.0194 & 0.0437 & 0.0269 \\
\bottomrule
\end{tabular}
\label{tab:exp-ablation}
\vspace{-5mm}
\end{wraptable}

\subsection{Ablation Study (RQ2)}
In this section, we analyze the contribution of each component in
\methodshort{} through ablation studies on the Clothing-Sports dataset. 
The results are reported in Table~\ref{tab:exp-ablation}, and the key findings are as follows:

(1) \underline{{\textbf{\textit{w/o} HDQ.}}}
Removing the hierarchical domain-aware quantizer leads to a performance drop. 
This result indicates that combining globally shared coarse-level quantization with adaptively routed fine-level quantization is more effective than the corresponding ablated tokenizer. The performance comparison supports the hierarchical design but does not by itself establish representation disentanglement.

(2) \underline{{\textbf{\textit{w/o} DAS MoE.}}}
Replacing the DAS MoE with standard dense FFN also degrades performance, suggesting that the combination of shared and sparsely activated specialized capacity is beneficial under unified cross-domain training.

(3) \underline{{\textbf{\textit{w/o} CRCL.}}}
Removing the Cross-Granularity Routing Consistency Learning module also leads to noticeable performance degradation. 
 This finding indicates that regularizing intra-item routing consistency is beneficial for recommendation performance.

\subsection{Tokenizer Design Analysis (RQ2)}  
Beyond the component-level ablations, we further investigate the design choices of the hierarchical domain-aware tokenizer. Specifically, we compare HD-REC with variants that remove shared codebooks, move the adaptive codebooks to the first layer, replace learned routing with random routing, or use a single expanded codebook at the final layer. The results are reported in Table~\ref{tab:exp-ablation-tokenizer}.

\begin{wraptable}{r}{0.48\textwidth}
\vspace{-5mm}
\centering
\footnotesize
\setlength{\tabcolsep}{2.2pt}
\caption{Tokenizer ablation results.}
\vspace{-2mm}
\begin{tabular}{lcccc}
\toprule
\multirow{2}{*}{\textbf{Model}} & \multicolumn{2}{c}{\textbf{Cloth.}} & \multicolumn{2}{c}{\textbf{Sport.}} \\
\cmidrule{2-5}
& \textbf{H@10} & \textbf{N@10} & \textbf{H@10} & \textbf{N@10} \\
\midrule
\cellcolor{gray!20}\textbf{Full HDQ} & \cellcolor{gray!20}\textbf{0.0256} & \cellcolor{gray!20}\textbf{0.0191} & \cellcolor{gray!20}\textbf{0.0418} & \cellcolor{gray!20}\textbf{0.0259} \\
\textit{w/} All Shared & 0.0239 & 0.0177 & 0.0391 & 0.0257 \\
\textit{w/} Specific First & 0.0233 & 0.0166 & 0.0389 & 0.0253 \\
\textit{w/o} Routing & 0.0238 & 0.0169 & 0.0385 & 0.0251 \\
\textit{w/} Expand Last & 0.0244 & 0.0179 & 0.0409 & 0.0257 \\
\bottomrule
\end{tabular}
\label{tab:exp-ablation-tokenizer}
\vspace{-5mm}
\end{wraptable}
(1) \underline{\textbf{\textit{w/} All Shared.}}
We replace HDQ with a standard RQ-VAE tokenizer where all codebooks are shared across domains. 
This variant replaces the adaptively routed final-layer codebooks with a shared codebook. Its lower performance indicates that additional input-dependent fine-level quantization capacity is beneficial for modeling heterogeneous item representations.

(2) \underline{\textbf{\textit{w/} Specific First.}}
Moving the adaptive codebooks to the first layer also degrades performance. This result is consistent with our design motivation that globally shared codebooks are suitable for coarse semantic encoding, while adaptive capacity is more useful for residual refinement.

(3) \underline{\textbf{\textit{w/o} Routing.}}
 We further investigate the importance of learned routing mechanism by replacing it with random routing, where each sample arbitrarily selects a codebook pathway. This leads to significant performance degradation, confirming the importance of data-dependent codebook selection.

 (4) \underline{\textbf{\textit{w/} Expand Last.}}
Expanding a single shared final-layer codebook does not match the performance of using multiple adaptively routed codebooks. This indicates that the gain does not result solely from increasing final-layer vocabulary capacity; the routing structure itself contributes to performance.

Overall, the results support the proposed hierarchical tokenizer design: shared codebooks are more suitable for coarse-grained semantic decomposition, whereas adaptive routing provides additional capacity for fine-grained domain-dependent information.

\subsection{Hyper-parameter Analysis (RQ3)}  
\begin{figure}[!t]
\centering
\includegraphics[width=1\linewidth]{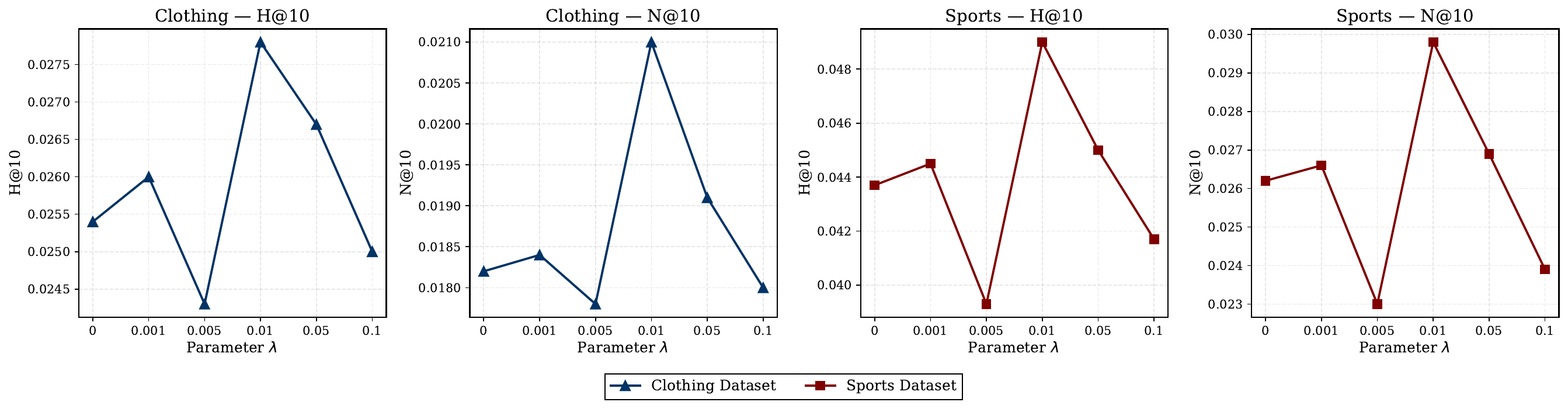}
\caption{The results of hyper-parameter experiments on the Clothing and Sports dataset.
}
\label{fig:exp-hyper}
\vspace{-4mm}
\end{figure}

To answer \textbf{RQ3}, we study the impact of the Cross-Granularity Routing Consistency Learning (CRCL) loss weight $\lambda_1$. Specifically, we vary $\lambda_1$ within $\{0, 0.001, 0.005, 0.01, 0.05, 0.1\}$ while keeping other hyper-parameters fixed. The performance trends are illustrated in Figure~\ref{fig:exp-hyper}.  
As $\lambda_1$ increases, the performance exhibits a clear \textit{inverted-U} trend. The best performance is achieved at $\lambda_1=0.01$ across both datasets, with relative improvements of 9.4\%/15.0\% on Clothing and 11.8\%/13.4\% on Sports (H@10/N@10) compared with the variant where $\lambda_1=0$. 

When $\lambda_1$ exceeds 0.01, the performance gradually degrades, indicating that excessive consistency regularization overly constrains token-level routing flexibility. Thus, we set $\lambda_1=0.01$ as the default to balance routing coherence and semantic expressiveness.



\subsection{Efficiency Analysis (RQ4)}

In terms of \textbf{RQ4}, we evaluate the inference efficiency of our proposed \methodshort{}. As shown in the infer time columns of Table~\ref{tab:exp-efficiency-consistency}, \methodshort{} introduces only a marginal inference overhead compared with the baseline TIGER. Specifically, \methodshort{} takes 106.2 ms and 107.8 ms per test instance on Clothing and Sports, respectively, which is only 2.6 ms and 2.1 ms slower than the baseline. This corresponds to a relative overhead of merely 2.5\% and 2.0\%, indicating that the additional inference cost brought by \methodshort{} is negligible in practice.



\subsection{Consistency Study (RQ5)}

\begin{wraptable}{t}{0.51\textwidth}
\centering
\vspace{-5mm}
\caption{Efficiency and routing-consistency comparison under beam size 200. Lower is better. Baseline denotes TIGER in the efficiency experiment and \textit{w/o} CRCL in the routing-consistency experiment.}
\scriptsize
\begin{tabular}{l|cc|cc}
\toprule
\multirow{2}{*}{\textbf{Model}} 
& \multicolumn{2}{c|}{\textbf{Infer Time (ms)}} 
& \multicolumn{2}{c}{\textbf{RV} ($\times10^{-4}$)} \\
\cmidrule{2-5}
& \textbf{Cloth.} & \textbf{Sport.} 
& \textbf{Cloth.} & \textbf{Sport.} \\
\midrule
\cellcolor{gray!20}\textbf{\methodshort{}} 
& \cellcolor{gray!20}106.2 
& \cellcolor{gray!20}107.8 
& \cellcolor{gray!20}0.429 
& \cellcolor{gray!20}0.589 \\
Baseline 
& 103.6 
& 105.7 
& 2.626 
& 3.348 \\
\bottomrule
\end{tabular}
\vspace{-5mm}
\label{tab:exp-efficiency-consistency}
\end{wraptable}

To evaluate item-level routing consistency of the MoE router, we introduce a metric called \emph{Routing Variance (RV)}.
For each token $t$ in an item, the router outputs a probability distribution over experts $\mathbf{p}_t \in \mathbb{R}^{E}$, where $E$ is the number of experts.
We compute the variance of routing probabilities across tokens belonging to the same item. For item $i$ with token set $T_i$, the routing variance is defined as:
\begin{equation}
\mathrm{RV}_i \;=\; \frac{1}{E}\sum_{e=1}^{E}\mathrm{Var}_{t\in \mathcal{T}_i}\!\left(p_{t,e}\right),
\end{equation}
where $\mathcal{T}_i$ denotes the set of tokens belonging to item $i$ and $p_{t,e}$ denotes the routing probability of token $t$ assigned to expert $e$.
The global RV is obtained by averaging across all items:
\begin{equation}
\mathrm{RV} \;=\; \mathbb{E}_{i}\!\left[\mathrm{RV}_i\right].
\end{equation}
A smaller RV indicates that tokens within the same item tend to be routed to similar experts, reflecting stronger routing consistency. As shown in Table~\ref{tab:exp-efficiency-consistency} right,
\methodshort{} achieves an approximately $6.0\times$ reduction in RV compared with the baseline, indicating that the proposed consistency regularizer substantially improves intra-item routing coherence.





%% file: 5RelatedWork.tex
\section{Related Works}

\noindent \textbf{Cross-domain Sequential Recommendation}.


Cross-domain recommender systems (CDR) transfer knowledge across domains to alleviate data sparsity. Early approaches relied on overlapping entities as bridges, such as EMCDR~\cite{cao2022cross} learning latent mappings via shared users, CoNet~\cite{hu2018conet} enabling bidirectional knowledge transfer through cross-connected neural networks, and HeroGRAPH~\cite{cui2020herograph} constructing shared heterogeneous graph to propagate information across domains. Recently, LLMs have advanced CDR by capturing semantic relationships without overlapping entities. LLM4CDSR~\cite{liu2025bridge} extracts cross-domain preferences from textual descriptions, and LeCDSR~\cite{wang2025lecdsr} generates unified semantic representations to support zero-shot transfer. However, effectively integrating diverse semantics while preserving specific knowledge remains a key challenge.

\noindent \textbf{Generative Recommendation Models}.
Generative recommendation replaces the traditional retrieval-and-ranking pipeline by directly generating the next item in a single stage. Typically, items are quantized into hierarchical semantic IDs and predicted sequentially using Transformer-based models. Existing research mainly focuses on improving quantization to preserve item semantics (e.g., TIGER, LETTER)~\cite{rajput2023recommender,wang2024learnableitemtokenizationgenerative} and optimizing architectures and training strategies (e.g., HSTU, OneRec)~\cite{zhai2024actions,deng2025onerec}. However, most existing works target single-domain datasets, and their effectiveness in cross-domain scenarios remains limited.

%% file: 6Conclusion.tex
\section{Conclusion}
In this paper, we presented \methodshort{}, a generative framework for pairwise cross-domain
sequential recommendation. \methodshort{} combines hierarchical domain-aware
quantization, domain-adaptive sparse experts, and cross-granularity
routing consistency learning. The framework uses globally shared
coarse-level codebooks together with adaptively routed fine-level
codebooks, and combines shared model capacity with sparsely activated
specialized capacity. Experiments on three public benchmark pairs show
consistent improvements over competitive baselines. Ablation studies
verify the contribution of each component, and routing analysis confirms
that CRCL improves intra-item routing consistency. Future work will
investigate direct measures of representation specialization and unified
training over a larger number of domains.

%% file: 7Limitation.tex
\section{Limitations}
Although \methodshort{} achieves consistent improvements on three public cross-domain recommendation datasets, several limitations remain. First, our experiments mainly focus on two-domain cross-domain sequential recommendation settings. While these datasets cover different application scenarios, the scalability and stability of \methodshort{} under more complex multi-domain settings with a larger number of heterogeneous domains require further investigation. Second, \methodshort{} relies on content-derived item representations and semantic ID tokenization. Therefore, its performance may be affected when item metadata is noisy, incomplete, or weakly aligned with users' actual preference signals. Third, our evaluation is conducted on standard offline benchmark datasets, while real-world recommendation systems often involve dynamic item catalogs, evolving user interests, changing domain distributions, and online feedback loops. The behavior of \methodshort{} under these more realistic deployment conditions has not been fully discussed in this work. In future work, we also plan to explore broader practical issues such as fairness and privacy preservation in generative cross-domain recommendation.

%% file: main.bib
@article{ma2024triple,
  title={Triple sequence learning for cross-domain recommendation},
  author={Ma, Haokai and Xie, Ruobing and Meng, Lei and Chen, Xin and Zhang, Xu and Lin, Leyu and Zhou, Jie},
  journal={ACM Transactions on Information Systems},
  volume={42},
  number={4},
  pages={1--29},
  year={2024},
  publisher={ACM New York, NY}
}

@inproceedings{cui2020herograph,
  title={HeroGRAPH: A Heterogeneous Graph Framework for Multi-Target Cross-Domain Recommendation.},
  author={Cui, Qiang and Wei, Tao and Zhang, Yafeng and Zhang, Qing},
  booktitle={ORSUM@ RecSys},
  year={2020}
}

@inproceedings{hu2018conet,
  title={Conet: Collaborative cross networks for cross-domain recommendation},
  author={Hu, Guangneng and Zhang, Yu and Yang, Qiang},
  booktitle={Proceedings of the 27th ACM international conference on information and knowledge management},
  pages={667--676},
  year={2018}
}

@inproceedings{cao2022cross,
  title={Cross-domain recommendation to cold-start users via variational information bottleneck},
  author={Cao, Jiangxia and Sheng, Jiawei and Cong, Xin and Liu, Tingwen and Wang, Bin},
  booktitle={2022 IEEE 38th International Conference on data engineering (ICDE)},
  pages={2209--2223},
  year={2022},
  organization={IEEE}
}

@article{jang2016categorical,
  title={Categorical reparameterization with gumbel-softmax},
  author={Jang, Eric and Gu, Shixiang and Poole, Ben},
  journal={arXiv preprint arXiv:1611.01144},
  year={2016}
}

@inproceedings{lee2022autoregressive,
  title={Autoregressive image generation using residual quantization},
  author={Lee, Doyup and Kim, Chiheon and Kim, Saehoon and Cho, Minsu and Han, Wook-Shin},
  booktitle={Proceedings of the IEEE/CVF conference on computer vision and pattern recognition},
  pages={11523--11532},
  year={2022}
}

@article{raffel2020exploring,
  title={Exploring the limits of transfer learning with a unified text-to-text transformer},
  author={Raffel, Colin and Shazeer, Noam and Roberts, Adam and Lee, Katherine and Narang, Sharan and Matena, Michael and Zhou, Yanqi and Li, Wei and Liu, Peter J},
  journal={Journal of machine learning research},
  volume={21},
  number={140},
  pages={1--67},
  year={2020}
}

@inproceedings{hu2025ids,
  title={From ids to semantics: A generative framework for cross-domain recommendation with adaptive semantic tokenization},
  author={Hu, Peiyu and Lu, Wayne and Wang, Jia},
  booktitle={Proceedings of the AAAI Conference on Artificial Intelligence},
  pages={14874--14882},
  year={2026}
}

@article{wang2025lecdsr,
  title={LeCDSR: Large language model enhanced cross-domain sequential recommendation},
  author={Wang, Shuliang and Zhu, Jiabao and Wang, Kaibo and Ruan, Sijie},
  journal={Information Fusion},
  pages={103762},
  year={2025},
  publisher={Elsevier}
}

@inproceedings{liu2025bridge,
  title={Bridge the domains: Large language models enhanced cross-domain sequential recommendation},
  author={Liu, Qidong and Zhao, Xiangyu and Wang, Yejing and Zhang, Zijian and Zhong, Howard and Chen, Chong and Li, Xiang and Huang, Wei and Tian, Feng},
  booktitle={Proceedings of the 48th International ACM SIGIR Conference on Research and Development in Information Retrieval},
  pages={1582--1592},
  year={2025}
}

@inproceedings{wang2024learnableitemtokenizationgenerative,
  title = {Learnable Item Tokenization for Generative Recommendation},
  author = {Wang, Wenjie and Bao, Honghui and Lin, Xinyu and Zhang, Jizhi and Li, Yongqi and Feng, Fuli and Ng, See-Kiong and Chua, Tat-Seng},
  booktitle = {International Conference on Information and Knowledge Management},
  year = {2024}
}

@article{rajput2023recommender,
  title={Recommender systems with generative retrieval},
  author={Rajput, Shashank and Mehta, Nikhil and Singh, Anima and Hulikal Keshavan, Raghunandan and Vu, Trung and Heldt, Lukasz and Hong, Lichan and Tay, Yi and Tran, Vinh and Samost, Jonah and others},
  journal={Advances in Neural Information Processing Systems},
  volume={36},
  pages={10299--10315},
  year={2023}
}

@article{hu2022lora,
  title={Lora: Low-rank adaptation of large language models.},
  author={Hu, Edward J and Shen, Yelong and Wallis, Phillip and Allen-Zhu, Zeyuan and Li, Yuanzhi and Wang, Shean and Wang, Lu and Chen, Weizhu and others},
  journal={ICLR},
  volume={1},
  number={2},
  pages={3},
  year={2022}
}

@article{deng2025onerec,
  title={Onerec: Unifying retrieve and rank with generative recommender and iterative preference alignment},
  author={Deng, Jiaxin and Wang, Shiyao and Cai, Kuo and Ren, Lejian and Hu, Qigen and Ding, Weifeng and Luo, Qiang and Zhou, Guorui},
  journal={arXiv preprint arXiv:2502.18965},
  year={2025}
}

@inproceedings{cao2022contrastive_,
  title={Contrastive cross-domain sequential recommendation},
  author={Cao, Jiangxia and Cong, Xin and Sheng, Jiawei and Liu, Tingwen and Wang, Bin},
  booktitle={Proceedings of the 31st ACM International Conference on Information \& Knowledge Management},
  pages={138--147},
  year={2022}
}

@article{zhai2024actions,
  title={Actions speak louder than words: Trillion-parameter sequential transducers for generative recommendations},
  author={Zhai, Jiaqi and Liao, Lucy and Liu, Xing and Wang, Yueming and Li, Rui and Cao, Xuan and Gao, Leon and Gong, Zhaojie and Gu, Fangda and He, Michael and others},
  journal={arXiv preprint arXiv:2402.17152},
  year={2024}
}

@inproceedings{hou2022towards,
  title={Towards universal sequence representation learning for recommender systems},
  author={Hou, Yupeng and Mu, Shanlei and Zhao, Wayne Xin and Li, Yaliang and Ding, Bolin and Wen, Ji-Rong},
  booktitle={Proceedings of the 28th ACM SIGKDD conference on knowledge discovery and data mining},
  pages={585--593},
  year={2022}
}

@inproceedings{hou2023learning,
  title={Learning vector-quantized item representation for transferable sequential recommenders},
  author={Hou, Yupeng and He, Zhankui and McAuley, Julian and Zhao, Wayne Xin},
  booktitle={Proceedings of the ACM Web Conference 2023},
  pages={1162--1171},
  year={2023}
}

@inproceedings{fan2022sequential,
  title={Sequential recommendation via stochastic self-attention},
  author={Fan, Ziwei and Liu, Zhiwei and Wang, Yu and Wang, Alice and Nazari, Zahra and Zheng, Lei and Peng, Hao and Yu, Philip S},
  booktitle={Proceedings of the ACM web conference 2022},
  pages={2036--2047},
  year={2022}
}

@article{dai2024deepseekmoe,
  title={Deepseekmoe: Towards ultimate expert specialization in mixture-of-experts language models},
  author={Dai, Damai and Deng, Chengqi and Zhao, Chenggang and Xu, RX and Gao, Huazuo and Chen, Deli and Li, Jiashi and Zeng, Wangding and Yu, Xingkai and Wu, Yu and others},
  journal={arXiv preprint arXiv:2401.06066},
  year={2024}
}

@article{deldjoo2024recommendation,
  title={Recommendation with generative models},
  author={Deldjoo, Yashar and He, Zhankui and McAuley, Julian and Korikov, Anton and Sanner, Scott and Ramisa, Arnau and Vidal, Rene and Sathiamoorthy, Maheswaran and Kasrizadeh, Atoosa and Milano, Silvia and others},
  journal={arXiv preprint arXiv:2409.15173},
  year={2024}
}

@article{chen2023cdr,
  title={Cdr-adapter: Learning adapters to dig out more transferring ability for cross-domain recommendation models},
  author={Chen, Yanyu and Yao, Yao and Chan, Wai Kin Victor and Xiao, Li and Zhang, Kai and Zhang, Liang and Ye, Yun},
  journal={arXiv preprint arXiv:2311.02398},
  year={2023}
}

@article{lee2025gram,
  title={GRAM: Generative Recommendation via Semantic-aware Multi-granular Late Fusion},
  author={Lee, Sunkyung and Choi, Minjin and Choi, Eunseong and Kim, Hye-young and Lee, Jongwuk},
  journal={arXiv preprint arXiv:2506.01673},
  year={2025}
}

@article{jin2025generative,
  title={Generative Multi-Target Cross-Domain Recommendation},
  author={Jin, Jinqiu and Zhang, Yang and Feng, Fuli and He, Xiangnan},
  journal={arXiv preprint arXiv:2507.12871},
  year={2025}
}

@article{su2024cross,
  title={Cross-domain recommendation via dual adversarial adaptation},
  author={Su, Hongzu and Li, Jingjing and Du, Zhekai and Zhu, Lei and Lu, Ke and Shen, Heng Tao},
  journal={ACM Transactions on Information Systems},
  volume={42},
  number={3},
  pages={1--26},
  year={2024},
  publisher={ACM New York, NY}
}

@inproceedings{geng2022recommendation,
  title={Recommendation as language processing (rlp): A unified pretrain, personalized prompt \& predict paradigm (p5)},
  author={Geng, Shijie and Liu, Shuchang and Fu, Zuohui and Ge, Yingqiang and Zhang, Yongfeng},
  booktitle={Proceedings of the 16th ACM conference on recommender systems},
  pages={299--315},
  year={2022}
}

@article{zhang2022cross,
  title={Cross-domain collaborative recommendation without overlapping entities based on domain adaptation},
  author={Zhang, Hongwei and Kong, Xiangwei and Zhang, Yujia},
  journal={Multimedia Systems},
  volume={28},
  number={5},
  pages={1621--1637},
  year={2022},
  publisher={Springer}
}

@article{zang2022survey_,
  title={A survey on cross-domain recommendation: taxonomies, methods, and future directions},
  author={Zang, Tianzi and Zhu, Yanmin and Liu, Haobing and Zhang, Ruohan and Yu, Jiadi},
  journal={ACM Transactions on Information Systems},
  volume={41},
  number={2},
  pages={1--39},
  year={2022},
  publisher={ACM New York, NY}
}

@inproceedings{li2011cross,
  title={Cross-domain collaborative filtering: A brief survey},
  author={Li, Bin},
  booktitle={2011 IEEE 23rd International Conference on Tools with Artificial Intelligence},
  pages={1085--1086},
  year={2011},
  organization={IEEE}
}

@article{zhu2021cross,
  title={Cross-domain recommendation: challenges, progress, and prospects},
  author={Zhu, Feng and Wang, Yan and Chen, Chaochao and Zhou, Jun and Li, Longfei and Liu, Guanfeng},
  journal={arXiv preprint arXiv:2103.01696},
  year={2021}
}

@article{li2025li_survey,
  title={A survey of generative recommendation from a tri-decoupled perspective: Tokenization, architecture, and optimization},
  author={Li, Xiaopeng and Chen, Bo and She, Junda and Cao, Shiteng and Wang, You and Jia, Qinlin and He, Haiying and Zhou, Zheli and Liu, Zhao and Liu, Ji and others},
  year={2025},
  journal={Preprints},
  publisher={Preprints}
}

@inproceedings{sun2019bert4rec,
  title={BERT4Rec: Sequential recommendation with bidirectional encoder representations from transformer},
  author={Sun, Fei and Liu, Jun and Wu, Jian and Pei, Changhua and Lin, Xiao and Ou, Wenwu and Jiang, Peng},
  booktitle={Proceedings of the 28th ACM international conference on information and knowledge management},
  pages={1441--1450},
  year={2019}
}

@inproceedings{kang2018self,
  title={Self-attentive sequential recommendation},
  author={Kang, Wang-Cheng and McAuley, Julian},
  booktitle={2018 IEEE international conference on data mining (ICDM)},
  pages={197--206},
  year={2018},
  organization={IEEE}
}

@inproceedings{kenton2019bert,
  title={Bert: Pre-training of deep bidirectional transformers for language understanding},
  author={Devlin, Jacob and Chang, Ming-Wei and Lee, Kenton and Toutanova, Kristina},
  booktitle={Proceedings of the 2019 conference of the North American chapter of the association for computational linguistics: human language technologies, volume 1 (long and short papers)},
  pages={4171--4186},
  year={2019}
}
